# Quasi-potential landscape in complex multi-stable systems

**Joseph Xu Zhou**[1,2,5*], **M. D. S. Aliyu**[3], **Erik Aurell**[4,5], **Sui Huang**[1,2,5*]

[1] Institute for Systems Biology, Seattle, WA, USA

[2] Institute for Biocomplexity and Informatics, University of Calgary, Calgary, Alberta, Canada;

[3] Mathematics and Statistics Department, University of Calgary, Calgary, Alberta, Canada;

[4] Department of Computational Biology, KTH, Stockholm, Sweden

[5] Kavli Institute for Theoretical Physics China, CAS, Beijing, China

*Authors for correspondence (Joseph.Zhou@systemsbiology.org; Sui.Huang@systemsbiology.org)

**Abstract**
Developmental dynamics of multicellular organism is a process that takes place in a multi-stable system in which each attractor state represents a cell type and attractor transitions correspond to cell differentiation paths. This new understanding has revived the idea of a quasi-potential landscape, first proposed by Waddington as a metaphor. To describe development one is interested in the "relative stabilities" of N attractors (N>2). Existing theories of state transition between local minima on some potential landscape deal with the exit in the transition between a pair attractor but do not offer the notion of a global potential function that relate more than two attractors to each other. Several *ad hoc* methods have been used in systems biology to compute a landscape in non-gradient systems, such as gene regulatory networks. Here we present an overview of the currently available methods, discuss their limitations and propose a new decomposition of vector fields that permit the computation of a quasi-potential function that is equivalent to the Freidlin-Wentzell potential but is not limited to two attractors. Several examples of decomposition are given and the significance of such a quasi-potential function is discussed.

**Notation**
**x**, $x$  - Variables of a dynamic system **x** = $(x_1, x_2, …)$
**x**\*     - stable steady states
**F**(**x**) - Driving forces of a dynamical system
$U(\mathbf{x})$ - Quasi-potential function
   $U^L$    - Lyapunov function
   $U^{int}$ - Proposed quasi-potential function based on integration
   $U^{prob}$ - Proposed quasi-potential function based on the probability density function of steady states
   $U^{norm}$ - Proposed quasi-potential function based on the normal decomposition
   $U^{ST}$ -   Proposed quasi-potential function based on the symmetric and antisymmetric decomposition
$V(x)$   Quasi-potential based on the least action function in the Freidlin-Wentzell theory

## 1. INTRODUCTION: The need for a quasi-potential function

For high-dimensional, non-linear dynamical systems that exhibit a large number of stable steady states (attractors) far from thermodynamic equilibrium, a characteristic system behavior of interest is manifest at the scale of transitions between these attractors across large distances in phase space. Such systems are epitomized by gene regulatory networks of *N* genes, where attractor states represent the cell types in the metazoan body. Accordingly, transitions between attractor states correspond to cell phenotype changes during normal development, homeostasis and disease [1,2]. But standard analysis of dynamical systems focuses on the existence and local properties of a given steady state (linear stability). Thus, it fails to capture the essence of complex systems that operate at time scales where the transitions between attractors constitute the characteristic behavior.

Specifically, in a system with *M* attractor states $\mathbf{x}_1^*, \mathbf{x}_2^*, \ldots \mathbf{x}_M^*$, one would like to obtain a sense of the "relative depth" [3] or more precisely, the ordering of the metastable attractor states through some energy-like function $U_1$, $U_2$, $U_3$… $U_M$. Such a function $U(\mathbf{x})$ for any point $\mathbf{x}$ in the state space of the system would inform about the probability and direction of transitions between attractor states in a noisy or perturbed system, for instance in understanding spontaneous cell fate choice (from among a set of alternative nearby attractors for a given initial state) or the feasibility of artificial reprogramming between particular cell types. Such ordering would also explain hierarchies and the arrow of time of ontogenesis.

Let us consider a deterministic system (network) of *N* variables $x_i$ (e.g. the activity of interacting genes) whose values describes the cell state $x(t) = (x_1(t), x_2(t), \cdots, x_N(t))^T$ and whose dynamics results from how each gene influences the activity of other genes (as invariantly pre-determined by the gene regulatory network that is hard-coded in the genome). Such dynamics is described by the first-order ordinary differential equations (ODEs) which in general are non-linear:

$$\frac{dx_1}{dt} = F_1(x_1, x_2, \cdots, x_N)$$
$$\frac{dx_2}{dt} = F_2(x_1, x_2, \cdots, x_N) \quad (1)$$
$$\vdots$$

or in vector form:

$$d\mathbf{x}/dt = \mathbf{F}(\mathbf{x})$$

Such a system represents an *influence network* in which the variables (network nodes) influence each other's values according to the network rules defined by the vector $\mathbf{F}(x)$. Thus, $\mathbf{F}(x)$ represent the "forces" acting to change the system state $x(x_1, x_2, \cdots, x_N)$ in this inertia-free system. In one-dimensional systems one always obtains a potential function $U^{int}(\mathbf{x})$ by integration which is a measure of the metastability of two (attractor) states $x_A$ and $x_B$ in a system:

$$U_{AB}^{int}(x) = U^{int}(x_B) - U^{int}(x_A) = \int_{x_A}^{x_B} F(x) dx \quad (2)$$

Equivalently, the driving force is proportional to the gradient of this potential function:

$$F(x) \sim -\frac{dU^{int}}{dx} \qquad (3)$$

In systems at thermodynamic equilibrium where the steady-state probability density function over **x** (giving the presence of thermal fluctuations) satisfies the Boltzmann distribution, $U^{int}$ is the potential energy and $U_{ab}^{int}$ is the path-independent potential difference between the two states $x_A$ and $x_B$. It is related to the transition probability via the Arrhenius equation:

$$P_{x_A \to x_B} = e^{-U_{AB}^{int}/\varepsilon^2} \qquad (4)$$

here $\varepsilon$ is the magnitude of noise [4].

Most works refer to transitions between equilibrium states of thermodynamic systems or simple cases such as Markovian one-dimensional non-equilibrium systems. Here we are interested in the rates of transition between states in far-from-equilibrium, non-conservative and high-dimensional open systems with a driving force emanating from the internal interactions as described by Eq. (1). For such cases, even if we have a gradient system where a potential function $U$ can be obtained by integration, i.e. where there exists a function $U^{int}(x)$ with the following properties:

$$\frac{\partial U^{int}}{\partial x_1} = -F_1(\boldsymbol{x}), \quad \cdots, \quad \frac{\partial U^{int}}{\partial x_i} = -F_i(\boldsymbol{x}), \quad \cdots, \quad \frac{\partial U^{int}}{\partial x_n} = -F_n(\boldsymbol{x}) \qquad (5)$$

such that one can obtain, by integration, a potential function $U^{int}(x)$

$$U^{int}(\boldsymbol{x}) = -\int F_1(\boldsymbol{x})dx_1 + \cdots + F_i(\boldsymbol{x})dx_i + \ldots + F_N(\boldsymbol{x})dx_n \qquad (6)$$

even then the transition rate for $x_A \to x_B$ is not path-independent and is determined by the potentials of attractor states and saddle points between them. The transition rate $P_{x_A \to x_B}$ is related to $U_{AB}^{int}(x) = U^{int}(x_S) - U^{int}(x_A)$ and the transition follows the *least action path* [5] as discussed below. Here $x_S$ is the saddle point between two attractors $\mathbf{x}_A$ and $\mathbf{x}_B$, as shown in Fig. 1.

Yet, a potential function in non-gradient systems which is loosely referred to as "quasi-potential" (indicated here by $\widetilde{U}$ when its computation is not further specified) is not without significance if properly defined such that it can serve as a quantity for the metastability ordering of (metastable) attractor states of the system in Eq. 1.

Specifically, for $\widetilde{U}$ to be of meaning for our purposes in characterizing cell state transitions in development we require that $\widetilde{U}$
(*i*) satisfies $\frac{d\widetilde{U}}{dt} < 0$ for $x \neq x^*$ and $\frac{d\widetilde{U}}{dt} = 0 \; for \; x = x^*$ (where $x^*$ are the stable steady states, which can be either fixed points or limit cycle) according to the stability theory of Lyapunov [6];
(*ii*) is related to the Freidlin-Wentzell action along the least action path between two states and hence permits the computation of the transition rate between them [5];

(iii) We also propose that the following transitivity is satisfied: if $\tilde{U}_a > \tilde{U}_b$ and $\tilde{U}_b > \tilde{U}_c$ then $\tilde{U}_a > \tilde{U}_c$ for any three states $x_a, x_b, x_c$.

Note that (i) is the stability property of an attractor state in a non-local sense; (ii) refers to the relationship of any two points and only (iii) pertains to multiple attractor states. This last requirement (iii) is only conjectured to be satisfied in a subset of systems and will be discussed in the discussion section. This last condition must be satisfied for $\tilde{U}$ to permit global ordering of metastability as described above.

## 2. THE PHYSICS PROBLEM: THE QUASI-POTENTIAL FUNCTION

Since high-dimensional, non-equilibrium systems generally are not gradient systems, i.e. Eq. (6) is not satisfied,

$$\frac{d\mathbf{x}}{dt} = \mathbf{F}(\mathbf{x}) \neq -\nabla \tilde{U} \tag{7}$$

By contrast, one can enforce a partial notion of a quasi-potential $\tilde{U}$, if we write the driving force as a sum of two terms:

$$\frac{d\mathbf{x}}{dt} = \mathbf{F}(\mathbf{x}) = -\nabla \tilde{U} + \mathbf{F}_r \tag{8}$$

Where $\mathbf{F}_r$ is the "remainder" beyond the component of the driving force that follows the potential gradients. Thus, we decompose the non-gradient vector field $\mathbf{F}(\mathbf{x})$, which is finite and smooth (at least twice differentiable), into two components: one that is the gradient of some "potential" $\tilde{U}$ and a second that represents the remainder of driving forces of the dynamic system. The question then is: what is the physical meaning of these terms? Given that there are infinite ways of decomposing a vector field into a sum of two fields, uniqueness of decomposition is imparted by the introduction of constraints which in turn embody the physical significance. Our objective here is to find decomposition such that the quasi-potential $\tilde{U}$ represents exactly the transition barriers associated with the state transitions among steady states and the "remainder" of the driving force $\mathbf{F}_r$ will not contribute to the efforts needed for the transition.

Little work has been done to the construction of quasi-potentials given a system $\frac{d\mathbf{x}}{dt} = \mathbf{F}(\mathbf{x})$ to determine the metastability ordering as defined above. Prigogine proposed a measure of global stability in nonlinear dynamic systems far from equilibrium to compare transition barriers between any pairs of attractors in a multistable system following a larger perturbation [7]. Efimov studied the mathematical properties of Lyapunov functions of multi-stable nonlinear systems but did not give the general method how to construct it [8]. Bhattacharya etc. proposed a simple method to visualize a landscape in lieu of a vector field. But it represents neither Lyapunov stability nor transition rates among different attractors[9]. Ao suggested a general method to construct the quasi-potentials using symmetric and antisymmetric decomposition[10–14]. The method of Ao can be shown to be mathematically equivalent to our approach under certain conditions but it is not convenient for computing the transition rate between attractors in general. Xing generalized Ao's work to prove that this approach is a mapping of a non-equilibrium dynamical system onto a Hamiltonian one [15] but misrepresented Ao's method as a generalized Helmholtz decomposition (see below). Using stochastic systems where the dynamics manifests as the temporal evolution of the probability density function and is described by a Fokker-Plank equation, Wang[16] championed the idea of a global potential function for gene networks using a decomposition that is similar but again, not identical to the Helmholtz decomposition (see below). Berglund reviews the Kramer's Law and the computation of the transition rate in multistable dynamic system. He gives a rigorous mathematical discussion of the limits of applying Kramer's Law in non-gradient system [17].

Here we discuss several *ad hoc* methods used to construct or visualize quasi-potential landscape and analyze whether they satisfy our first two conditions: i) global metastability and ii) agreement with the Freidlin-Wentzell theory for calculating the transition rates. Then we present the 'normal decomposition' as a systematic method that yields a quasi-potential that meets the above conditions. Mathematical derivation and several examples are given to demonstrate the validity and the utility of our new approach.

### 3. COMPARISON OF VECTOR FIELD DECOMPOSITION

We first review several commonly used methods of vector field decompositions and the meaning of the components.

### 3a. The Helmholtz decomposition

The *Hodge decomposition theorem* states that any sufficiently smooth, rapidly decaying vector field in *n*-dimensional space can be uniquely decomposed into the sum of a curl-free potential field $U^H(\mathbf{x}, t)$ and a divergence-free curl term [18]. Helmholtz's theorem is a special case of Hodge decomposition theorem in 3D, which can be written in the following equations:

$$\mathbf{F}(\mathbf{x}) = -\nabla U^H(\mathbf{x}) + \mathbf{F_c}(\mathbf{x})$$

$$\mathbf{F_c}(\mathbf{x}) = \nabla \times \mathbf{A}(\mathbf{x}) \tag{9}$$

$$\nabla \cdot \mathbf{F_c}(\mathbf{x}) = 0$$

The potential $U^H(x, t)$ can be found by solving the following Poisson equations:

$$\nabla \cdot F(x) = -\nabla \cdot \nabla U^H(\mathbf{x}) + \nabla \cdot (\nabla \times A(\mathbf{x})) \tag{10}$$

Since the curl term is divergence free, i.e. $(\nabla \times A(x)) = 0$, we have a Poisson equation:

$$\nabla^2 U^H(\mathbf{x}) = -\nabla \cdot \mathbf{F}(\mathbf{x}) \tag{11}$$

Without boundary condition, the decomposition is not unique because any specific solution which satisfies Eq. (12) plus any harmonic function will also be qualified. But the decomposition is unique with the boundary conditions that $U^H(x, t) \to 0$ when $x$ approaches infinity. Therefore, the potential $U^H(x, t)$ can be directly "read off" from the dynamic equations (1) without time stepping process by solving the static Poisson equations with known right-hand terms and boundary conditions.

The meaning of this decomposition is that any dynamical field can be divided into two parts: a conservative irrotational field whose gradient connects the attractors or repellors (analogous to sinks and sources in the field theory) and a solenoidal curl field that is not influenced by sinks or sources. However, given that the divergence-free curl field has been used to explain the driving force of the gradient-free limit cycle [19], it is noteworthy that the curl field does not necessarily consist of closed trajectories. It can be made of open trajectories with varying direction, as shown in Fig. 2. Nevertheless, this decomposition permits the study of the factors that influence the period of limit cycles (the curl term) and their stability (the gradient term) separately.

We can now determine if the potential function obtained from the Helmholtz decomposition, $U^H$ satisfies the Lyapunov criterion for stability, our requirement (i), i.e. in particular, if $\frac{dU^H}{dt} < 0$ for $x \neq x^*$. For a two-dimensional system taking the time derivative of $U^H$ from Eq. (9), we obtain:

$$\frac{dU^H}{dt} = -\left(\frac{\partial U^H}{\partial x}\right)^2 - \left(\frac{\partial U^H}{\partial y}\right)^2 + (\nabla U^H, \mathbf{F_c}) \tag{12}$$

Since we only know that $\nabla \cdot \mathbf{F_c}(\mathbf{x}) = 0$, $(\nabla U^H, \mathbf{F_c})$ can be either positive or negative. Thus, $\frac{dU^H}{dt}$ is not necessarily non-positive, and hence Lyapunov stability is not guaranteed in this decomposition. Therefore, $U^H$ does not satisfy our criterion (i) and it is not a measure of global metastability.

### 3b. Decomposition based on the flux of the probability, *U~* -Ln*P*

An often taken, *ad hoc* approach to obtain a quasi-potential function $\widetilde{U}$ and to visualize it as landscape[16,20,21] stems from the intuitive notion that in a probabilistic form of the system, the more stable states (lower potential $\widetilde{U}$) are more probable in the state space. Thus, the steady state probability for each state $x$ is then linked to a potential $U^{prob}$ according to the Boltzmann law:

$$U^{\text{prob}}(x, t) \sim -\ln(P(x, t)) \tag{13}$$

Again, since in general $\mathbf{F}(x) \neq \nabla U^{\text{prob}}$, the driving force $\mathbf{F}(x)$ requires an additional component besides the gradient term. With the constraints of Eq. (13), i.e. a gradient force emanating from a probability flux as one component, Wang and colleagues derived the following decomposition [22]:

The time evolution of the probability density function $P(x, t)$ is governed by the Fokker-Plank Equation, which can accurately describe the transition dynamics in a multistable dynamic system if the noise is Gaussian distribution:

$$\frac{\partial P(\mathbf{x},t)}{\partial t} = -\nabla \cdot (\mathbf{F}\, P) + D \cdot \nabla^2 P \tag{14}$$

in which the drift term is the driving force $\mathbf{F}(\mathbf{x})$ of our system. Note that diffusion term $D$ needs to be large enough for the probability density function $P(\mathbf{x},t)$ to achieve a good coverage of whole state space. The temporal change of the probability density can be written in terms of the probability flux $\mathbf{J}(\mathbf{x},t)$:

$$\frac{\partial P(\mathbf{x},t)}{\partial t} = -\nabla \cdot \mathbf{J}(\mathbf{x},t)$$

$$\mathbf{J}(\mathbf{x},t) = F\,P - D \cdot \nabla P \tag{15}$$

Since at steady states $\frac{\partial P(\mathbf{x},t)}{\partial t} = 0$, the divergence of the steady state probability flux $\mathbf{J}_{ss}$ vanishes, i.e. $\nabla \cdot \mathbf{J}_{ss}(\mathbf{x},t) = 0$, we obtain:

$$\begin{aligned}
\mathbf{F} &= \frac{\mathbf{J}_{ss}}{P_{ss}} + D \cdot \frac{\nabla P_{ss}}{P_{ss}} \\
&= -\nabla - \ln(P_{ss}) \cdot D + \frac{\mathbf{J}_{ss}}{P_{ss}} \\
&= -\nabla U^{\text{prob}} + \mathbf{F}_c
\end{aligned} \tag{16}$$

In this vector field decomposition based on steady state probability density, the driving force $\mathbf{F}(\mathbf{x})$ has been decomposed into two terms:

$$-\nabla U^{\text{prob}} = -\nabla(-\ln(P_{ss})) \cdot D \text{ and}$$

$$\mathbf{F}_c = \frac{\mathbf{J}_{ss}}{P_{ss}} \tag{17}$$

Thus, the first term consists indeed of a gradient of a potential $U^{prop}$. The second term is the flux $\frac{\mathbf{J}_{ss}}{P_{ss}}$ and has a curl nature. However, it is not a curl in the same sense as in the Helmholtz decomposition since generally it is not divergence free $\nabla \cdot \frac{\mathbf{J}_{ss}}{P_{ss}} \neq 0$. Only $\mathbf{J}_{ss}$ itself is divergence-free. Thus this decomposition is not a Helmholtz decomposition as claimed elsewhere [21].

To test if $U^{prob}$ of this decomposition meets both our requirement (i) and (ii), we insert $P = e^{-U^{\text{prob}}/D}$ and $F = -\nabla U^{\text{prob}} + \mathbf{F}_c$ into the Fokker-Plank Equation:

$$\frac{\partial P}{\partial t} = -\nabla(\mathbf{F} \cdot P) + D \cdot \nabla^2 P \tag{18}$$

$$\frac{\partial}{\partial t}\left(e^{-U^{\text{prob}}/D}\right) = -\nabla\left((-\nabla U^{\text{prob}} + \mathbf{F}_c)\cdot e^{-U^{\text{prob}}/D}\right) + D\cdot\nabla^2 e^{-U^{\text{prob}}/D}$$

and obtain

$$(-\nabla U^{\text{prob}}, \mathbf{F}_c) = D\,\nabla\cdot\mathbf{F}_c \tag{19}$$

That is, the two vector fields are not perpendicular unless $\mathbf{F}_c = \frac{J_{ss}}{P_{ss}}$ is divergence free – which is not necessary so as argued above. In other words, this decomposition, while motivated by $U^{\text{prob}} \sim -\ln(P)$, is actually a field decomposition $\mathbf{F} = -\nabla U + \mathbf{F}_c$ with the constraint $(-\nabla U^{\text{prob}}, \mathbf{F}_c) = D\,\nabla\cdot\mathbf{F}_c$. We can now again determine if the potential function obtained, $U^{\text{prob}} \sim -\ln(P_{ss})$ satisfies the Lyapunov criterion for stability, our requirement (i), i.e. in particular, if $\frac{dU^{\text{prob}}}{dt} < 0$ for $x \neq x^*$

For a two-dimensional system taking the time derivative of $U^{\text{prob}}$ we obtain:

$$\begin{aligned}\frac{dU^{\text{prob}}}{dt} &= -\left(\frac{\partial U^{\text{prob}}}{\partial x}\right)^2 - \left(\frac{\partial U^{\text{prob}}}{\partial y}\right)^2 + (\nabla U^{\text{prob}}, \mathbf{F}_c) \\ &= -\left(\frac{\partial U^{\text{prob}}}{\partial x}\right)^2 - \left(\frac{\partial U^{\text{prob}}}{\partial y}\right)^2 + D\,\nabla\cdot\mathbf{F}_c\end{aligned} \tag{20}$$

Thus, $\frac{dU^{\text{prob}}}{dt}$ is not necessarily non-positive, and hence Lyapunov stability is not guaranteed in this decomposition. Therefore, generally speaking, $U^{\text{prob}} \sim -\ln(P_{ss})$ cannot be considered a measure of global metastability. One argument to save $U^{\text{prob}} \sim -\ln(P)$ is that Eq. (19) will hold if noise term $D$ goes to zero in the limit, i.e. if $D \to 0$, then $D\,\nabla\cdot\mathbf{F}_c \to 0$. It implies that the gradient component and the remainder $\mathbf{F}_c$ would be perpendicular to each other and that the Lyapunov criterion is also satisfied. However, the noise term $D$ cannot be too small. If $D$ indeed goes to zero, state transitions between attractors will never happen and $P_{ss}$ will not be a global probability distribution but only a probability distribution in local attractors as the result of the balance between noise and driving forces.

We next examine the meaning for the case when the constraint is such that the gradient force and the remainder force are perpendicular to each other because this would guarantee satisfaction of the Lyapunov stability criterion.

### 3c. The normal decomposition
From the discussion above, the constraint that the gradient term is perpendicular to the "remainder" force $\mathbf{F}_\perp$ is of interest:

$$\mathbf{F} = -\nabla U^{norm} + \mathbf{F}_\perp \tag{21}$$

Taking the same approach as in Eq. 20, for a two-dimensional system, we obtain the time derivative for $U^{norm}$. Supposing that we have a Lyapunov function for a two-dimensional system, its time derivative along any trajectory driven by the dynamic system in Eq. (1) is:

$$\begin{aligned}
\frac{dU^{norm}}{dt} &= \frac{\partial U^{norm}}{\partial x}\dot{x} + \frac{\partial U^{norm}}{\partial y}\dot{y} \\
&= \frac{\partial U^{norm}}{\partial x}\left(-\frac{\partial U^{norm}}{\partial x} + F_\perp^x\right) + \frac{\partial U^{norm}}{\partial y}\left(-\frac{\partial U^L}{\partial y} + F_\perp^y\right) \\
&= -\left(\frac{\partial U^{norm}}{\partial x}\right)^2 - \left(\frac{\partial U^{norm}}{\partial y}\right)^2 + (\nabla U^L, \mathbf{F}_\perp)
\end{aligned} \quad (22)$$

If the gradient of $U^{norm}$ is normal to the remainder force $F_\perp$, i.e.

$$(\nabla U^{norm}, \mathbf{F}_\perp) = 0 \quad (23)$$

then the function $U^{norm}$ will decrease monotonically with time during the process of reaching the steady states, thus satisfying Lyapunov's condition for global metastability:

$$\frac{dU^{norm}}{dt} = -\left(\frac{\partial U^{norm}}{\partial x}\right)^2 - \left(\frac{\partial U^{norm}}{\partial y}\right)^2 < 0 \quad (24)$$

We can therefore see that the condition for the normal decomposition has an obvious physical meaning, in that for $(\nabla U^{norm}, \mathbf{F}_\perp) = 0$, $U^{norm}$ corresponds to a Lyapunov function $U^{norm}$ of dynamical systems and can represent the global metastability. Thus, we seek a decomposition of a sufficiently smooth vector field into a conservative potential field $U^{norm}$ and the remaining forces $\mathbf{F}_\perp$,

$$\begin{aligned}
\mathbf{F} &= -\nabla U^{norm} + \mathbf{F}_\perp \\
(-\nabla U^{norm}, \mathbf{F}_\perp) &= 0
\end{aligned} \quad (25)$$

The physical interpretation is that for a ball in an attractor state that is perturbed to exit with least "energy" against the system's "driving force" $\mathbf{F}(x)$ that keeps it in the attractor, we can decompose the field such that $\mathbf{F}_\perp$ will not contribute to this process but only a gradient field $U^{norm}$ contributes. Based on Freidlin-Wentzell's large deviation theory of a stochastic process discussed below [5], $U^{norm}$ can thus be used to compute the transition rate in non-equilibrium dynamic system.

The normal potential $U^{norm}$ can also be read off the systems equations without time stepping solution. We can calculate the potential field $U^{norm}$ as follows:

$$(\nabla U^{norm}, \mathbf{F}_\perp + \nabla U^{norm}) = 0 \quad (26)$$

This equation is the Hamilton-Jacobi equation (HJE) which can be written in a component format:

$$\frac{\partial U^{norm}}{\partial x_1} \cdot \left(F_{x_1} + \frac{\partial U^{norm}}{\partial x_1}\right) + \cdots + \frac{\partial U^{norm}}{\partial x_i} \cdot \left(F_{x_i} + \frac{\partial U^{norm}}{\partial x_i}\right) + \cdots + \frac{\partial U^{norm}}{\partial x_n} \cdot \left(F_{x_n} + \frac{\partial U^{norm}}{\partial x_n}\right) = 0 \quad (27)$$

The HJE provides for any nonequilibrium dynamic systems a general mathematical framework to construct a quasi-potential function which allows us to compare the relative stability of different attractors. This extends the global stability analysis of Lyapunov's theory to multi-stable dynamic systems. The quasi-potential function $U^{norm}$ also epitomizes the epigenetic landscape which could explain the driving force behind cell differentiation in multi-cellular organism, which can be used to calculate the transition rates between different cellular attractors (more details in Section 4)

Note that existence of a solution for Eq. (27) is by no means guaranteed for all dynamic systems. However, here we only deal with a subset of problems encountered in biology rather than a general dynamic system. We now propose that the dynamical systems which represent organismal development admit such decomposition. Further research is needed to find the conditions for the existence of such decomposition and verify them in the real biological systems.

It is far from trivial to solve nonlinear PDE like HJE which is still a topic of active research in applied mathematics. Since there usually exist no analytical solutions, numerical methods with careful convergence and stability analysis have to be applied to solve HJE in the real biological systems. A numerical method, the Newton-Raphson method, is used in this paper, which is described in the Supplement.

**3d. The relationship between the normal decomposition and the symmetric-antisymmetric decomposition**

The symmetric-antisymmetric decomposition proposed by Ao is under certain conditions mathematically equivalent to the normal decomposition approach presented above[23].

**The symmetric-antisymmetric decomposition when $D = I$**

For a dynamic system with governing equations in Eq. (1), the symmetric-antisymmetric decomposition can be achieved as follows[14] :

$$(S + T) \cdot F = -\nabla U^{ST}$$
$$(S - T)D(S + T) = S \qquad (28)$$

$U^{ST}$ is the quasi-potential derived from the symmetric-antisymmetric decomposition. $S$ is a symmetric operator, $T$ is an antisymmetric operator. Since it is usually difficult to find the analytical solutions for $S$ and $T$, $S$ *can be solved as a* symmetric matrix while $T$ as an antisymmetric matrix in finite dimensional approximation. $D$ is a diffusion matrix. Eq. (1) can be rewritten as the following:

$$F = -\nabla U^{ST} + F_r = (S + T) \cdot F + F_r \qquad (29)$$

Then $F_r$ can be written as

$$F_r = (I - S - T) \cdot F \qquad (30)$$

$I$ is an identity matrix.

To follow our perpendicular decomposition, the inner product of two terms is:

$$\begin{aligned}(\nabla U^{ST}, F_r) &= ((S+T)\cdot F, (I-S-T)\cdot F) \\ &= F^T(S+T)^T(I-(S+T))F \\ &= F^T(S-(S-T)(S+T))F\end{aligned} \quad (31)$$

If $D$ is the identify matrix, i.e. $D = I$, then from Eq. (28), we have

$$(S-T)(S+T) = S \quad (32)$$

And from Eq. (31) and (32), it is easy to show that then $(\nabla U^{ST}, F_r) = 0$, i.e. the gradient of the quasi-potential $\nabla U^{ST}$ is perpendicular to $F_r$. Therefore, the symmetric-antisymmetric decomposition is mathematically identical to our normal decomposition if the diffusion matrix $D$ is set to be an identity matrix $I$.

**Symmetric-antisymmetric decomposition when $D \neq I$**

The ensuing question is what the quasi-potential $U^{ST}$ obtained from the symmetric-antisymmetric decomposition will be if $D \neq I$. We can again evaluate this using the two conditions which we proposed in Section 2. First, we determine of the $U^{ST}$ decrease monotonically with time during the approach to the steady states, thus satisfying Lyapunov's condition for global metastability:

$$\frac{dU^{ST}}{dt} = \nabla U^{ST}\, F = -F^T(S+T)F = -F^T S F < 0 \quad (33)$$

We can therefore see, if the solution for Eq. (28) exists, $U^{ST}$ is indeed a Lyapunov function of a dynamical system which can represent global metastability.

However, if $D$ is not an identity matrix $I$, i.e. $(\nabla U^{ST}, F_r) \neq 0$, the remainder $F_r$ is not perpendicular to $\nabla U^{ST}$ and thus will contribute to the state transition. . Then, based on Freidlin-Wentzell's large deviation theory of a stochastic process (discussed below) [5], $U^{ST}$ cannot be used to compute the transition rate in non-equilibrium dynamic system.

## 4. FREIDLIN-WENTZELL THEORY OF LARGE DEVIATION IN MULTI-STABLE SYSTEMS

### 4a. Background

For a dynamic system with deterministic forces $F(x, t)$,

$$\frac{dx0}{dt} = F(x0, t) \quad (34)$$

Now let us consider that the system is under a stochastic perturbation $\varepsilon$:

$$\frac{dx}{dt} = F(x, t) + \varepsilon \cdot \xi(t) \quad (35)$$

If $\varepsilon$ is sufficiently small, the perturbed system will converge to the original dynamical system, i.e. $\|x - x0\| \to 0$. However, if the perturbation is a random process with a small average amplitude but with occasional large excursion, the perturbed dynamic system will behave differently. Freidlin and

Wentzell [5] proposed a large deviation theory of stochastic process as a theoretical framework for the analysis of dynamical systems with multiple attractors. Supposing that a dynamical system satisfies the Langevin dynamics, the governing equations are described by the following ODEs:

$$\begin{cases} \dot{x}_1 = f_1(x_1, \ldots, x_n) + \xi_1(t) \\ \vdots \\ \dot{x}_i = f_i(x_1, \ldots, x_n) + \xi_i(t) \\ \vdots \\ \dot{x}_n = f_{n1}(x_1, \ldots, x_n) + \xi_n(t) \end{cases} \tag{36}$$

Supposing that a ball is perturbed to go from state $\mathbf{x}_A^* \to \mathbf{x}_B^*$ by a stochastic process, one defines an action function $V_{AB}$ to measure the "energy" barrier to be overcome for this transition:

$$V_{AB} = \frac{1}{2} \min\{ \int_{tA}^{tB} \left[ \sum_{i=1}^{n} \|\dot{x}_i - f_i(\mathbf{x})\|^2 \right] dt \} \tag{37}$$

Here the action function $V_{AB}$ is defined as a time integral of the square of the "remainder" of the dynamic equations (deviation from deterministic trajectory) over the whole trajectory $X(t)$ from attractor $\mathbf{x}_A$ to $\mathbf{x}_B$. If a ball is only driven by the "forces" specified in the deterministic part of the ODEs (28), it will correspond to a "free fall" in the "gravity field" and the action is zero. If the ball is perturbed against the "forces", the remainder term $\|\dot{x}_i - f_i(\vec{x})\|^2$ will not be zero. We integrate them over time along the whole trajectory and obtain the total action when a ball switches from attractor $\mathbf{x}_A$ to $\mathbf{x}_B$. Based on the *variation principle*, there exists a unique minimum action $V_{AB}$ which is an objective measure of the difficulty for a state transition in a non-equilibrium dynamic system.

**4b. The relationship between the Freidlin-Wentzell potential and normal decomposition**
Although the Freidlin-Wentzell potential $V$ and the normal potential $U^{norm}$ are defined in different ways, they are mathematically related. For a dynamic system, $\mathbf{F} = -\nabla U^{norm} + \mathbf{F}_\perp$, with $(-\nabla U^{norm}, \mathbf{F}_\perp) = 0$. We can rewrite the Wentzell potential as:

$$\begin{aligned} V_{AB} &= \frac{1}{2} \min\{ \int_{tA}^{tB} \left[ \|\dot{\mathbf{X}} - \mathbf{F}\|^2 \right] dt \} \\ &= \frac{1}{2} \min\{ \int_{tA}^{tB} \left[ \|\dot{\mathbf{X}} + \nabla U^{norm} - \mathbf{F}_\perp\|^2 \right] dt \} \\ &= \frac{1}{2} \min\{ \int_{tA}^{tB} \left[ \|\dot{\mathbf{X}} - \mathbf{F}_\perp - \nabla U^{norm}\|^2 \right] dt + 4 \int_{tA}^{tB} \left[ (\dot{\mathbf{X}} - \mathbf{F}_\perp) \cdot \nabla U^{norm} \right] dt \} \\ &= \frac{1}{2} \min\{ \int_{tA}^{tB} \left[ \|\dot{\mathbf{X}} - \mathbf{F}_\perp - \nabla U^{norm}\|^2 \right] dt + 4 \int_{tA}^{tB} \left[ \dot{\mathbf{X}} \cdot \nabla U^{norm} \right] dt \} \\ &= \frac{1}{2} \min\{ \int_{tA}^{tB} \left[ \|\dot{\mathbf{X}} - \mathbf{F}_\perp - \nabla U^{norm}\|^2 \right] dt + 4(U_B^{norm} - U_A^{norm}) \} \\ &= 2(U_B^{norm} - U_A^{norm}) \end{aligned} \tag{38}$$

During the process of exiting an attractor, the *least action path* of a ball follows the governing equation:

$$\mathbf{F} = \nabla U^{norm} + \mathbf{F}_\perp \tag{39}$$

Intuitively speaking, the *least action path* out of attractors will be driven by a landscape which reverses the original one, i.e. the hills are reverted to valleys and valleys are changed to be hills. However, the reversal only holds for the gradient component because the normal remainder $\mathbf{F}_\perp$ still retain the same direction as before [24]. As long as we are within the same attractor, the Freidlin-Wentzell potential is exactly twice as big as the potential from the normal decomposition, $V_{AB} = 2(U_B^{norm} - U_A^{norm})$.

When a "ball" transitions from one attractor to another, the Freidlin-Wentzell potential only accounts for the uphill "energy", which is two times that of $U^{norm}$. Once it goes over the "saddle" point (point $X_s$ in Fig. 1), the Wentzell potential $V$ is zero for the remaining free-fall path. The same applies for a ball that transitions through *many* attractors in between: The Wentzell potential is equal to the sum of all uphill potential between two points. All downhill paths contribute nothing to the Freidlin-Wentzell potential.

### 4c. Wentzell potential and spontaneous transition rate

For a system which is under a large deviation perturbation: $\frac{dx}{dt} = \mathbf{F}(\mathbf{x}, t) + \varepsilon \cdot \boldsymbol{\xi}(t)$, $V_{AB}$ as defined above Eq. (37), the probability of spontaneous transition $\mathbf{x}_A \to \mathbf{x}_B$ is [5]:

$$P_{x_A \to x_B} = e^{-\frac{V_{AB}}{\varepsilon^2}} \qquad (40)$$

Accordingly, for the transition between $x_B \to x_A$ the probability is:

$$P_{x_B \to x_A} = e^{-\frac{V_{BA}}{\varepsilon^2}} \qquad (41)$$

If $V_{AB} > V_{BA}$, under random perturbation, a ball will stay longer in state $\mathbf{x}_B$, i.e. state $\mathbf{x}_B$ is globally more stable than state $\mathbf{x}_A$, and thus, the directionality of the spontaneous transition points from state $\mathbf{x}_A$ to state $\mathbf{x}_B$.

### 4c. Wentzell potential and least action path of spontaneous transition

Although the Wentzell potential $V_{AB}$ can be mathematically expressed as the normal potential $U^{norm}$, it offers additional information which cannot be directly derived from $U^{norm}$: In determining $V_{AB}$, not only a minimum value is computed, but also a least action path $X_{AB}$ between two points is obtained. However, even if there is a least action path, a state transition $\mathbf{x}_A \to \mathbf{x}_B$ does not necessarily have to follow exactly this path only. Since the perturbation is random, the effective state transition can happen along any path that topologically connects these two states. The majority of probable paths fall near the area of the least action path, i.e. for any path $\mathbf{X_t}\ (\mathbf{x}_A \to \mathbf{x}_B)$, the least action path is $X_{AB}$ and

$$P(\|\mathbf{X_t} - \mathbf{X_{AB}}\| < \varepsilon) = e^{-\frac{V_{AB}}{\varepsilon^2}} \qquad (42)$$

Here, $\|\ \ \|$ is defined as a norm, $\|X\| = (\sum_{i=1}^n (x_i)^2)^{1/2}$. It means that the bigger the noise, the further away can the actual transition paths deviate from the least action path $\mathbf{X_{AB}}$.

## 5. EXAMPLES

In this section, two examples are presented to show which decomposition can give the correct transition rates between metastable attractors in nonlinear, high-dimensional dynamical systems. Quasi-potentials $U(\mathbf{x})$ are computed using different decompositions: The Helmholtz decomposition (section 3a), the $U\sim$ - Ln$P$ decomposition (section 3b) and the normal decomposition (section 3c). The transition barriers $\Delta U$ are calculated based on the quasi-potentials constructed from these decompositions. The results are compared with the Wentzell action functions which represent the correct transition rates among different attractors.

**Example 1. Spiral dynamic system with a single attractor**

A two-dimensional non-gradient dynamical system reveals the effects of the non-vanishing remainder component of the vector field. The governing equations are:

$$\begin{cases} \dot{x} = -x + y^2 \\ \dot{y} = -y - xy \end{cases} \tag{43}$$

The perpendicular decomposition of the vector field can be easily found (eq. 26, 27):

$$\begin{cases} \dot{x} = -\dfrac{\partial U^{\text{norm}}}{\partial x} + F_\perp^x = -\dfrac{\partial (x^2+y^2)/2}{\partial x} + y^2 \\ \dot{y} = -\dfrac{\partial U^{\text{norm}}}{\partial x} + F_\perp^y = -\dfrac{\partial (x^2+y^2)/2}{\partial y} - xy \end{cases} \tag{44}$$

The quasi-potential functions and the remainders from the perpendicular decomposition are:

$$U^{\text{norm}} = (x^2 + y^2)/2$$

$$\begin{cases} F_\perp^x = y^2 \\ F_\perp^y = -xy \end{cases} \tag{45}$$

It can be easily verified that:

$$\dfrac{\partial U^{\text{norm}}}{\partial x} \cdot F_\perp^x + \dfrac{\partial U^{\text{norm}}}{\partial y} \cdot F_\perp^y = x \cdot y^2 - y \cdot xy = 0 \tag{46}$$

The Helmholtz decomposition can be constructed in a straightforward manner:

$$\begin{cases} \dot{x} = -\dfrac{\partial U^{\text{H}}}{\partial x} + F_c^x = -\dfrac{\partial (x^2+y^2+xy^2)/2}{\partial x} + \dfrac{3}{2}y^2 \\ \dot{y} = -\dfrac{\partial U^{\text{H}}}{\partial x} + F_c^y = -\dfrac{\partial (x^2+y^2+xy^2)/2}{\partial y} \end{cases} \tag{47}$$

We have

$$U^{\text{H}} = (x^2 + y^2 + xy^2)/2$$

$$\begin{cases} F_c^x = \dfrac{3}{2}y^2 \\ F_c^y = 0 \end{cases} \tag{48}$$

As mentioned before, the Helmholtz decomposition is not unique. The superposition of the function above with any harmonic function still satisfies the Helmholtz decomposition. However, this example illustrates the general characteristics of Helmholtz decomposition: the remainder and the gradient components are not necessarily perpendicular, thus, according to section 4, it is not the right decomposition to give the correct transition rate.

Since $U^{prob}$ in the –lnP decomposition (3b) depends on solving the steady states of the Fokker-Planck equation, which usually does not have an analytical solution, here we calculate $U$~ -lnP using a numerical method (Finite Difference Method), see details in the Supplement. Different noise levels in Fokker-Planck Equation lead to different distributions $P(\mathbf{x})$ in steady states. $U^{prob}$ = -lnP*D were calculated with different noise level (D=0.015, 0.045, 0.09) to cover a wider range of results from –lnP decomposition.

The quasi-potentials at nine points (relative to the origin, i.e. $U(0,0) = 0$) from four different methods were calculated based on the methods described above and shown in Fig. 3. As can be recognized, the quasi-potential $U^H$ (in circle) from the Helmholtz decomposition significantly departs from the Wentzell potential and thus is not good for calculating the transition probability. The quasi-potential from the -lnP decomposition $U^{Prob}$ has two problems. First, if noise level $D$ is small (D=0.015), $U^{prob}$ (in symbol 'x') does not "cover" the peaks on both sides. Second, even if noise level $D$ is large enough to sample the whole landscape, its quasi-potential $U^{prob}$ (in pentagon and diamond) are not symmetrical due to the asymmetric probability flux in the phase space. Here only the quasi-potential derived from normal decomposition $U^{Norm}$ (in square) agrees with half of the Freidlin-Wentzell potential (in triangle). The numerical method for calculating the Freidlin-Wentzell potential is in the Supplement.

**Example 2. Two-dimensional dynamical system with four attractors**

Since our original motivation was to determine the relative meta-stability of attractors in a multi-stable dynamical system, an example with four attractors is chosen to illustrate the various quasi-potential functions that can be extracted from the decomposition methods and allow us to compare the transition rates and global ordering of attractors they predict. (The Helmholtz decomposition is excluded from the analysis for its obvious invalidity). Theoretical predictions from the two other decompositions are compared with the Freidlin-Wentzell action functions which can represent the transition rate between attractors caused by stochastic perturbation with large deviations.

The governing equations of our system are:

$$\begin{cases} \dfrac{dx}{dt} = F_1(x,y) = -1 + 9x - 2x^3 + 9y - 2y^3 \\ \dfrac{dy}{dt} = F_2(x,y) = 1 - 11x + 2x^3 + 11y - 2y^3 \end{cases} \quad (49)$$

Since the dynamical system has a symmetric structure, it can be decomposed into a gradient part and a perpendicular remainder.

$$\begin{cases} \dfrac{dx}{dt} = F_1(x,y) = -\dfrac{\partial U^{norm}}{\partial x} + F_\perp^x = (10x - 2x^3 - y - 1) + (10y - 2y^3 - x) \\ \dfrac{dy}{dt} = F_2(x,y) = -\dfrac{\partial U^{norm}}{\partial y} + F_\perp^y = (10y - 2y^3 - x) - (10x - 2x^3 - y - 1) \end{cases} \quad (50)$$

The quasi-potential function $U^{norm}$ from the normal decomposition is shown in Fig. 4 (A). It has four attractors designated as A, B, C and D, with the quasi-potential function $U^{norm}$ as follows:

$$U^{norm}(x,y) = -5(x^2+y^2) + \frac{1}{2}(x^4+y^4) + xy + x \tag{51}$$

We can now examine the ordering of the attractors with respect to their relative metastability. Using Eq. (51), we find: $U^{norm}(D) > U^{norm}(B) > U^{norm}(C) > U^{norm}(A)$. In this case we have a transitive set. i.e. if $U^{norm}(D) > U^{norm}(B)$, $U^{norm}(B) > U^{norm}(C)$, then $U^{norm}(D) > U^{norm}(C)$ and so forth. However, this does not need to be so in general (see discussion for more details).

We can verify that the quasi-potential function $U^{norm}(x,y)$ in Eq. (51) satisfies the Hamilton-Jacobi Equation Eq. (26). From the vector field plot in Fig. 4, it is clear that the gradient components contribute to the driving force toward the attractors (Fig. 4(B)) while the remaining perpendicular components capture the rotational movement in the phase space (Fig. 4(C)). However, since the rotational driving forces are perpendicular to the gradient components they do not contribute to the transition between the different attractors.

Using the same method as in the previous example, the decomposition for extracting $U^{prob}$ can be numerically computed with varying noise levels. In Fig. 4(D), the results for the quasi-potential $U^{prob}$ with various noise levels are plotted alongside $U^{norm}$ from the normal decomposition. As shown in previous example, the accuracy of $U^{prob}$ suffers if the noise level is too small to cover the whole landscape. In this case, we need $D = 20$ to get a reasonable accurate landscape.

Then we need to calculate Wentzell action function $V$ as the gold standard to validate two decompositions here. Because the Wentzell action function is usually too complicated to be found analytically, Eq. (37) can be re-written in discrete form to find an approximate solution (See Supplement for more details). We can calculate the least action path (LAP) for the attractor transition from $x_A$ to $x_D$ and verse versa, the attractor transition from $x_A$ to $x_C$ and verse versa(see Fig. 5). It is clear that the Wentzell action function applies during the "uphill" process while it is zero during the "downhill" process(see Fig. 5). The least action path from attractor from $x_D$ to $x_A$, which is different from the path from $x_A$ to $x_D$. Here the Wentzell action function is smaller than for the transition from $x_A$ to $x_D$ because the attractor state $x_D$ as a "higher elevation" than $x_A$ in the quasi-potential "landscape" $U^{norm}$. (see Fig. 5(A)(B)). Note that the least action path $\mathbf{x}_A \to \mathbf{x}_D$ is also different from the reverse path $\mathbf{x}_D \to \mathbf{x}_A$, and the least action path $\mathbf{x}_A \to \mathbf{x}_C$ is different from the reverse path $\mathbf{x}_C \to \mathbf{x}_A$(see Fig. 5). Such "path irreversibility" is a common characteristic of nonlinear dynamic systems.

Based on the Freidlin-Wentzell large deviation theory, the transition rate between two points can be determined from the transition barriers $\Delta U$ along the least action path (LAP). Instead of comparing the transition rates directly, $\Delta U$ calculated from the quasi-potential $U^{norm}$ and by $U^{prob}$ are compared with the Freidlin-Wentzell action functions $V$, as shown in Fig. 6. It is shown that $U^{norm}$ agree well with the half of the Freidlin-Wentzell action functions $V$ while $U^{prob}$ mostly do not agree with it. It could be argued that the accuracy of $U^{prob}$ could be further improved by using bigger noise and better numerical methods. However, from various numerical test, $U^{prob}$ does not significantly improve its accuracy more than the example presented here.

## 6. DISCUSSION

We show here that the quasi-potential $U^{norm}$ obtained from the normal decomposition has a non-local meaning, in that the sign of $\Delta U^{norm}$ represents the relative stability of a pair of adjacent attractors with respect to the spontaneity to transition from one attractor to another. The spontaneity implies that the system can, in the presence of noise in the state variables *x*, i.e., without explicit external driving force that directs this transition as epitomized by signaling molecules in cell biology, move from a given state to the most probable neighboring state. No matter what is the transition path between two neighboring fixed point attractors, $\mathbf{x}_A$ and $\mathbf{x}_B$, if $U_A^{norm} > U_B^{norm}$, it is always easier to transition from point $\mathbf{x}_A$ to point $\mathbf{x}_B$ and vice versa. By contrast, for the alternative decomposition based on $U^{prob}$~- lnP , this equivalency between potential and transition rate is not exact although for practical purposes, $U_S^{prob} - U_A^{prob}$ (where $U_S$ is the potential at the "saddle" (exit) point between these attractors) often scales with $P_{x_A \to x_B}$ [16].

For the transition $\mathbf{x}_A \to \mathbf{x}_C$ in cases where the attractors $\mathbf{x}_A$ and $\mathbf{x}_C$ are not adjacent, but separated by attractor B that needs to be traversed, i.e., $\mathbf{x}_A \to \mathbf{x}_B \to \mathbf{x}_C$ with "saddle" points $\mathbf{x}_{S_1}$ between $\mathbf{x}_A, \mathbf{x}_B$ and saddle points $\mathbf{x}_{S_2}$ between $\mathbf{x}_B, \mathbf{x}_C$ , the above approach can be expanded[25]**:**

$$\begin{aligned} P_{x_A \to x_C} &= P_{x_A \to x_B} \cdot P_{x_B \to x_C} = e^{-\frac{V_{AB}}{\varepsilon^2}} \cdot e^{-\frac{V_{BC}}{\varepsilon^2}} \\ &= e^{-\frac{U_{S_1}^{Norm} - U_A^{Norm}}{\varepsilon^2}} \cdot e^{-\frac{U_{S_2}^{Norm} - U_B^{Norm}}{\varepsilon^2}} \end{aligned} \quad (52)$$

The case with more than one attractor in between can also be calculated with the similar formula like Eq. (52).

We started with stating that the motivation for defining a quasi-potential function is the desire to compare the relative (meta)stability of attractor states in systems with multiple (more than two) attractors in order to have an universal tool to predict cell fate transitions and the "efforts" needed to reprogram cell types. However, existing theories of quasi-potential functions have been developed for the one-to-one relationship of two points (Freidlin-Wentzell theory, Kramer's Law). Two caveats are in order here regarding the interpretation of *ΔU*<sup>*norm*</sup> in systems with more than two attractors.

### 6a. *ΔU* and "fate choice" in multi-potent cells
First, the attractors with lower the quasi-potentials *U*<sup>*norm*</sup> have higher probability to be occupied when a system reaches its steady states - which may take a long time for a rugged landscape with small noise. This shall remind us that *Δ U*<sup>*norm*</sup> alone does not allow one to predict the "fate choice" of a stem cell in an attractor within a multi-attractor system that faces alternative transitions (representing its potential to differentiate into multiple distinct cell types spontaneously). The local landscape topography needs to be considered. Similar to the case for equilibrium systems, speaking of potentials, irrespective of whether path-dependent or not, one needs to bear in mind that *Δ U*<sup>*norm*</sup> between two points is not a general measure of spontaneity of a process but that the latter depends on the local kinetic barriers, as illustrated in Fig 1. Referring to the constellation in Fig. 1 it is clear that noise (or temperature) may be just high enough but not too high, such that one will find that the system in attractor A will prefer to move to B rather than to C (in the absence of catalysis). Only under sufficient noise (or high temperature) and time will the preference for C in steady state become manifest.

### 6b. A global potential landscape?

Second, we now return to the more profound motivation behind the interest for a global quasi-potential function: the assessment of the "relative (meta)stability" of attractor states, that is, their ordering with respect to spontaneous transitions into each other. Such a global behavior of a system, as opposed to local, peri-attractor behavior, is the scale of system change at which complex biological processes, such as development, takes place. But is such an ordering possible? Can we directly compare the values of the quasi-potentials $U_i^{norm}$ (*i*=1, 2, 3..*m*) for *m* attractors and extract meaningful information as to long-term fate choice – aside from the afore-mentioned issue of kinetic inhibition? This is tempting to ask because, unlike the Freidlin-Wentzell potential difference *ΔV* which is defined for pairs of states, we can compute from the system equations directly the potential value $U^{norm}$ (*S*) for any state *S*.

If $U^{norm}$ captures the global behavior one would postulate that our conjectured third condition, the transitivity of any three points *a, b, c* with respect to their quasi-potential *U* would be satisfied: If $U_a^{norm}$ > $U_b^{norm}$ and $U_b^{norm}$ > $U_c^{norm}$, then $U_a^{norm}$ > $U_c^{norm}$. The example 2 constitutes such a transitive system. However, there is no guarantee that transitivity is satisfied for all systems. Although $\Delta U^{norm}$ scales monotonically with ΔV (Wentzell action function) which is defined for points within the same basin of attraction, (most usefully, as difference between the attractor state and the exit state at the saddle) there is no simple mathematical relationship between $\Delta U^{norm}$ and ΔV along the transition trajectory between two attractors and is thus is "agnostic" for global ordering.

Only in those cases where there is transitivity among all the potential values of attractors *i* , $U_i^{norm}$ (consistency) in the directed graph in which the edge arrows depict the direction of higher transition probability $P(\mathbf{x}_i \rightarrow \mathbf{x}_j)$ between pairs of attractors $\mathbf{x}_i$ and $\mathbf{x}_j$ can we speak of a "global" landscape. This condition surely will not be satisfied for systems exhibiting biological phenomena such as circadian cycle, cell cycle and other biological limit cycles. However, in some class of systems as lucidly epitomized in the qualitative picture of Waddington's epigenetic landscape that describe the essential feature of cell development or in the fitness landscape of evolution, such global landscapes appear to adequately capture the developmental or evolutionary dynamics and constraints. In differentiation, it visualizes the barriers of trans-differentiation or reprogramming from one cell type to another. This suggests the possibility of a global ordering of (meta)stability for a particular yet to be defined subclass of networks to which our example 2 belong in which the global transitivity of *U* is satisfied. Thus it could be that evolved networks that govern acyclic phenotype transitions, such as development or evolution have for some unknown reasons a quasi-potential landscape that satisfies the transitivity condition (iii) – which is a subset of multi-attractor systems. The facility to compute for each attractor *i* of a biological network a quasi-potential $U_i$ opens the possibility to study this phenomenon.

## Acknowledgements
This work was supported by the Natural Sciences and Engineering Research Council of Canada (NSERC), Canadian Institutes of Health Research (CIHR) and Alberta Innovates.

## Figure Legends

**Figure 1**.  Schematic quasi-potential $U$ and the transitions among stable steady states (attractors) in a 1D multistable dynamical system.   The transition rate $P_{A \to B}$ is not determined by $\Delta U_{AB}$, but by $\Delta U_{AS_1}$. Similarly, the transition rate $P_{A \to C}$ is not determined by $\Delta U_{AC}$, but by $\Delta U_{AS_2}$.

**Figure2.**  Curl is not necessarily a driving force for a limit circle. The vector field of a divergence-free dynamical system has open trajectories.  The governing equations of the dynamical system are as follows:

$$\frac{dx}{dt} = -y; \frac{dy}{dt} = -x$$

**Figure 3** (A) Vector field of the driving forces of  the dynamic system in Example 1. Note how the non-vanishing remainder components $U_\perp$ cause the vector field to be non-symmetric even if the underlying quasi-potential $U^{\text{norm}}$  is symmetric. (B) The quasi-potential $U^{\text{norm}}$ reconstructed from the normal decomposition and the positions of nine points for the comparison in the next figure. (C) Different quasi-potentials constructed from the Helmholtz decomposition, the normal decomposition, -lnP decomposition and the Freidlin-Wentzell action function for the comparisons of nine positions (states) in phase space.  (Online version in colour.)

**Figure 4.** (A) Quasi-potential function $U^{norm}$ derived from the normal decomposition. The white circles represent attractors *A, B, C, D*; the grey circles denote the saddle points. (B) Vector field plot of the gradient component of the normal decomposition super-positioned on the contour plot of $U^{Norm}$;  (C) Vector field plot of the $\boldsymbol{F}_\perp$ , the remainder component of the normal decomposition super-positioned on the contour plot of $U^{Norm}$; (D) Quasi-potential $U^{\text{norm}}$ calculated from the normal decomposition and the $U^{\text{prob}}$ calculated from $-lnP$ decomposition with various noise levels ($D$=10, 15 and 20). (Online version in colour.)

**Figure 5.** (A) The least action path (LAP) for the attractor transition from $x_A$ to $x_D$. The white dot denotes the starting point; black dot is the end point. The attractors A, B, C, D are noted as in Fig. 4. $V(t)$ is the Wentzell action function at every time step, $V$ is the total Wentzell action function at each time $t$. (B) The least action path from attractor from $x_D$ to $x_A$, which is different from the path from $x_A$ to $x_D$. Here the Wentzell action function is smaller than for the transition from $x_A$ to $x_D$ because the attractor state $x_D$ as a "higher elevation" than $x_A$ in the quasi-potential "landscape" $U^{\text{norm}}$. (C) The least action path for the attractor transition from $x_A$ to $x_C$. It is clear that the Wentzell action function applies during the "uphill" process while it is zero during the "downhill" process; (D) The least action path for the attractor transition from $x_C$ to $x_A$. The Wentzell action function is smaller than for the transition from $x_A$ to $x_C$ because attractor $x_C$ is "higher" than $x_A$ in the quasi-potential $U^{\text{norm}}$ (Online version in colour.)

**Figure 6.** (A) Quasi-potential $U^{\text{prob}}$ at the attractors and "saddle" points derived from the -lnP decomposition with noise level *D*=20; (B) Quasi-potential $U^{\text{norm}}$ at the same points derived from the normal decomposition. Attractors *A, B, C, D* are denoted as in Fig. 4; (C) Potential barriers $\Delta U$ that needs to be overcome for the transition from each attractor to every other one. $\Delta U$ were calculated using the quasi-potential function from the normal decomposition $U^{\text{norm}}$ and -lnP decomposition $U^{\text{prob}}$ based on the topology of attractors and "saddle" points in between. Then $\Delta U$ were compared with the Freidlin-Wentzell action functions $V$ calculated from numeric minimization. (Online version in colour.)

## Short title for page headings
Multi-stable quasi-potential landscape

<END>

# Supplement information

### *Solving* Hamilton-Jacobi equation derived from *normal decomposition*

The Hamilton-Jacobi equation is a nonlinear partial differential equation, which usually has no analytical solutions. However, $U^{norm}$ can be solved numerically using the iterative Newton-Raphson method after boundary conditions are specified for real problems [1,2].

First, we discrete the quasi-potential $U(x)$ to finite dimension $\begin{pmatrix} U_1 \\ \vdots \\ U_n \end{pmatrix}$ and use differential matrix $[M_i] \begin{pmatrix} U_1 \\ \vdots \\ U_n \end{pmatrix}$ to replace the original differentiation operator $\frac{\partial U(x)}{\partial x_i}$, Eq. (27) can be written in a discrete form with some algebra transformations:

$$\Phi(\boldsymbol{U_k}(x)) = \sum_{i=1}^{n} \left( [M_i] \begin{Bmatrix} U_1 \\ \vdots \\ U_n \end{Bmatrix}_k \cdot \begin{Bmatrix} F_i(x_1) \\ \vdots \\ F_i(x_n) \end{Bmatrix} + [M_i] \begin{Bmatrix} U_1 \\ \vdots \\ U_n \end{Bmatrix}_k \cdot [M_i] \begin{Bmatrix} U_1 \\ \vdots \\ U_n \end{Bmatrix}_k \right) = 0 \quad (1)$$

We can calculate the Jacobian matrix:

$$\boldsymbol{J} = \frac{\partial \vec{F}}{\partial \vec{U}} = \sum_{i=1}^{n} \left( [M_i] \cdot \begin{Bmatrix} F_i(x_1) \\ \vdots \\ F_i(x_n) \end{Bmatrix} + 2[M_i] \begin{Bmatrix} U_1 \\ \vdots \\ U_n \end{Bmatrix}_k \cdot [M_i] \right) \quad (2)$$

Then we can solve the linear system for time step $k$:

$$\boldsymbol{J} \cdot \Delta \boldsymbol{U_k} = -\Phi(\boldsymbol{U_k}) \quad (3)$$

We derive the iterative formula:

$$\boldsymbol{U_{k+1}} = \boldsymbol{U_k} + \Delta \boldsymbol{U_k} \quad (4)$$

With an educated guess of initial value $\boldsymbol{U_0}$ and a convergence criteria $\|\boldsymbol{U_{k+1}} - \boldsymbol{U_k}\| < \varepsilon$, the quasi-potential $U(x)$ is solved in this iterative numerical scheme.

### *Solving U~-lnP with finite difference method*

We employ finite difference methods (FDM) to solve Fokker-Planck Equation numerically. For the simplification, a two-dimensional example is described here but the similar numerical scheme can be applied to *n*-dimensional problems. Here a square area is divided into square lattice boxes with a space $h$ and the mesh points $\big((x_{1,i}, x_{2,j}) = (i*h, j*h), i,j = 0,1,\ldots,N$ within the area.

The Fokker-Planck equation has first and second differentiation in space and first differentiation in time direction. Let $\tau$ denotes the time step, so that $t_k = k * \tau, t = 0,1,\ldots,M$. Then $P_{i,j}^k$ denotes the $P$ value on mesh point $(x_{1,i}, x_{2,j})$ at time $t_k$, and $F_{1,i,j}^k$ denotes the $F_1$ value on mesh point $(x_{1,i}, x_{2,j})$. We used a difference scheme that corresponds to Euler's scheme for the Fokker-Planck equation:

$$\frac{P_{i,j}^{k+1} - P_{i,j}^k}{\tau} = -P_{i,j}^k \left( \frac{F_{1,i+1,j}^k - F_{1,i-1,j}^k}{2h} + \frac{F_{2,i,j+1}^k - F_{2,i,j-1}^k}{2h} \right)$$

$$-F_{1,i,j}^k \frac{P_{i+1,j}^k - P_{i-1,j}^k}{2h} - F_{2,i,j}^k \frac{P_{i,j+1}^k - P_{i,j-1}^k}{2h}$$

$$+D\left( \frac{P_{i+1,j}^k - 2P_{i,j}^k + P_{i-1,j}^k}{h^2} + \frac{P_{i,j+1}^k - 2P_{i,j}^k + P_{i,j-1}^k}{h^2} \right) \quad (5)$$

$$= \Delta P(x_1, x_2, t)$$

Starting with the initial non-uniform conditions, we can step from any value of $t$ to $t + \tau$ with $P(x_1, x_2, t + \tau) = P(x_1, x_2, t) + \tau * \Delta P(x_1, x_2, t)$ for all of the mesh points $(x_1, x_2)$ in the area. The boundary conditions specify the values on the boundary. Here the Neumann or Dirichlet boundary conditions have no influence upon the result. More complicated methods, such as Runge-Kutta method can be used to improve the accuracy of time stepping. However, the basic formulas are the same as the one above.

**Solving the least-action trajectory in discrete form by conjugate gradient (CG) method**
Supposing that a dynamical system satisfies the Langevin dynamics, the governing equations are described by the following ODEs:

$$\begin{cases} \dot{x}_1 = f_1(x_1, \ldots, x_n) + \xi_1(t) \\ \vdots \\ \dot{x}_i = f_i(x_1, \ldots, x_n) + \xi_i(t) \\ \vdots \\ \dot{x}_n = f_{n1}(x_1, \ldots, x_n) + \xi_n(t) \end{cases} \quad (6)$$

Supposing that a ball is perturbed to go from state $\mathbf{x}_A^* \to \mathbf{x}_B^*$ by a stochastic process, one defines an action function $V_{AB}$ to measure the "energy" barrier to be overcome for this transition:

$$V_{AB} = \frac{1}{2}\min\{\int_{tA}^{tB}\left[\sum_{i=1}^{n}\|\dot{\mathbf{x}}_1 - f_i(\mathbf{x})\|^2\right]dt\} \quad (7)$$

Because the Wentzell action function is usually too complicated to be found analytically, Eq. (7) can be re-written in discrete form to find an approximate solution, as shown below:

$$V_{AB}(X(t)) = \Delta t \sum_{k=1}^{M-1}\sum_{i=1}^{n}\left[\left\|\frac{x_i^{k+1} - x_i^k}{\Delta t} - \frac{1}{2}(f_i^{k+1} - f_i^k)\right\|^2\right] \quad (8)$$

Here the total time over the trajectory $\mathbf{X}(t)$ is divided into $M-1$ equal time steps $\Delta t$. The time integral of action function is approximated with the sum of actions in each time step. The rate $\frac{dx}{dt}$ is approximated with the first-order difference equation $\frac{x_i^{k+1}-x_i^k}{\Delta t}$. To find the least-action trajectory in discrete form, initially two attractors are connected with a straight line and the conjugate gradient method (**CG**) is used to minimize the action function $V_{AB}(X(t))$ [3]. To ensure numerical stability, we started with 16 time segments and gradually increase it to be 32, 64, 128, …, until the change of the action function will be smaller than certain threshold $\epsilon$ (usually we set $\epsilon = 0.001$).

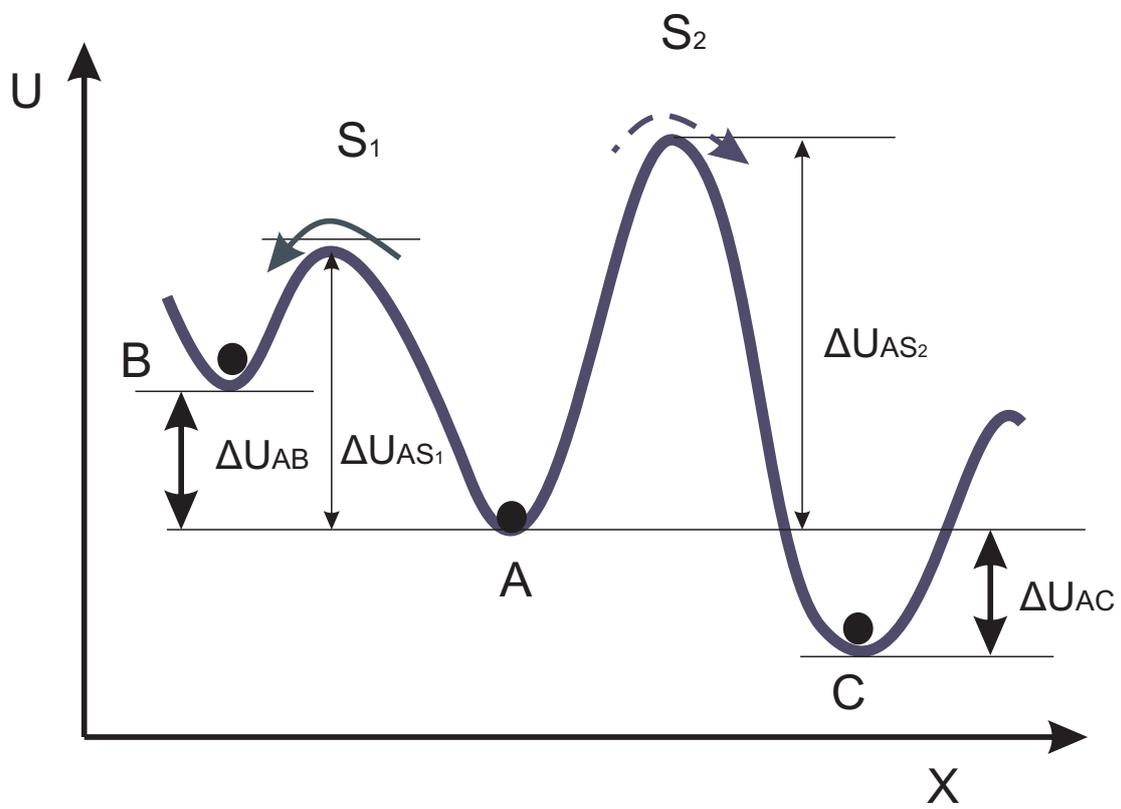

**Fig. 1**

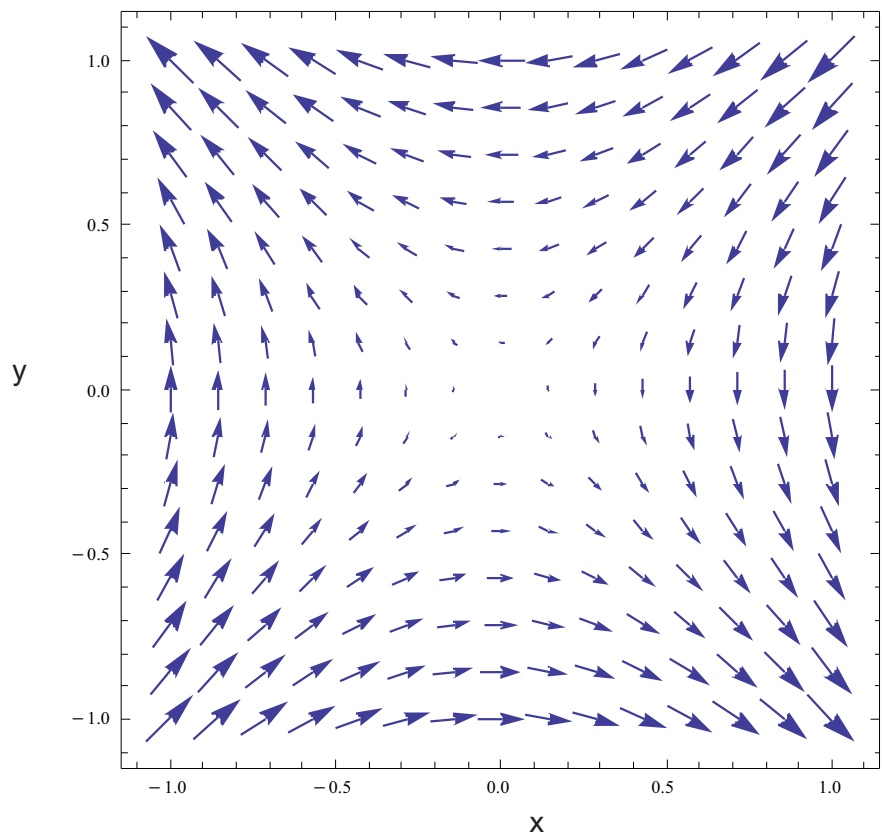

**Fig. 2**

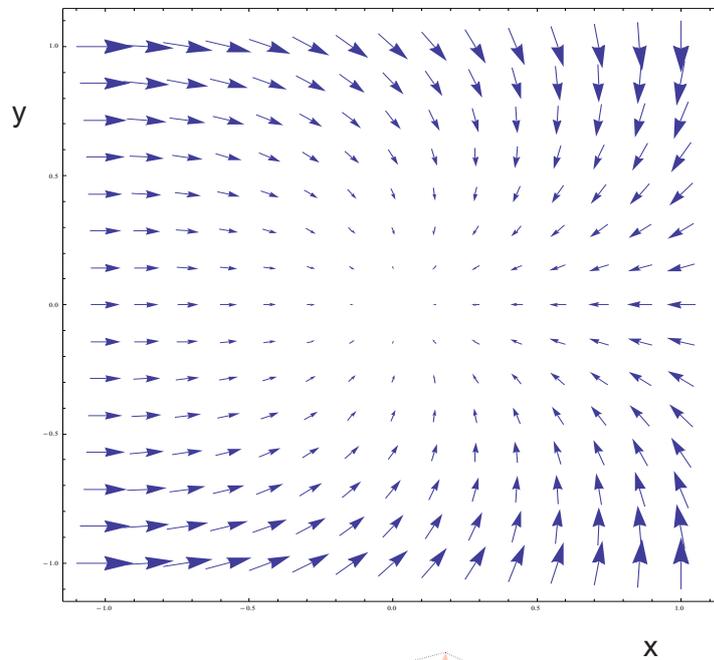

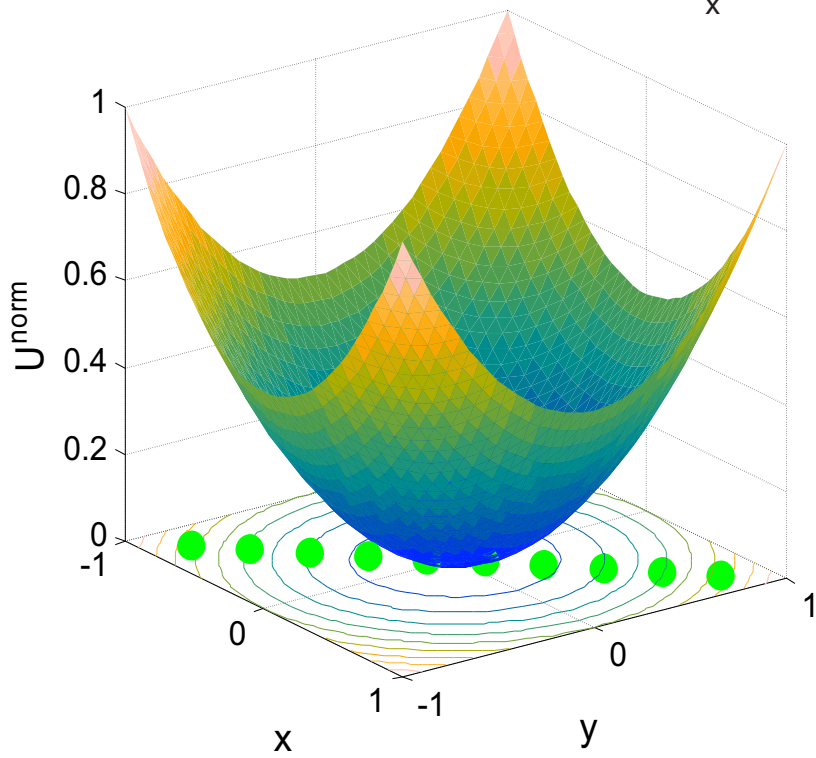

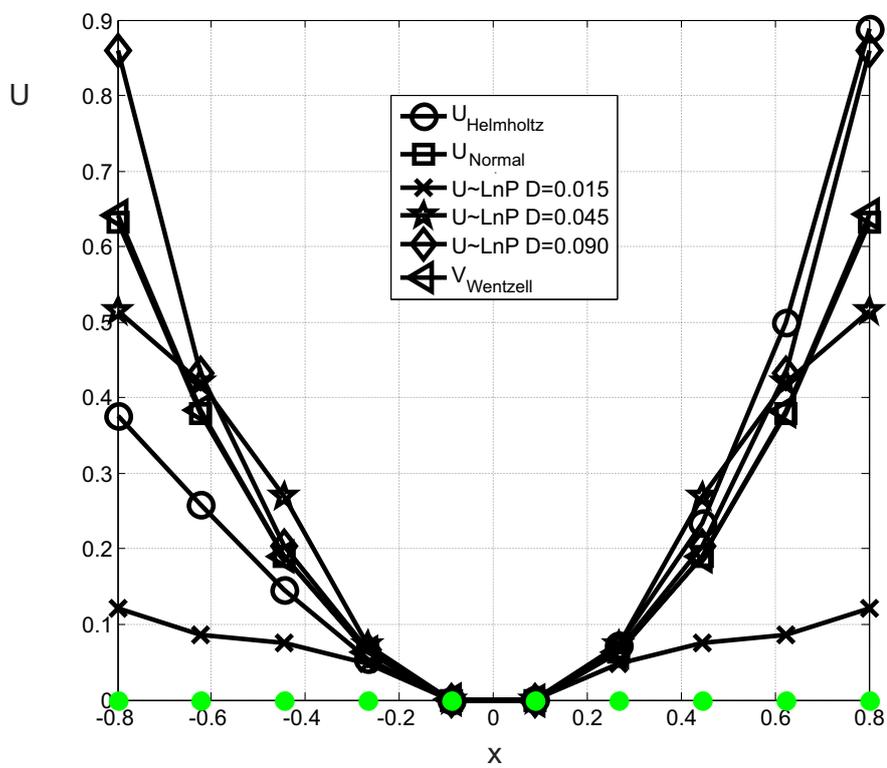

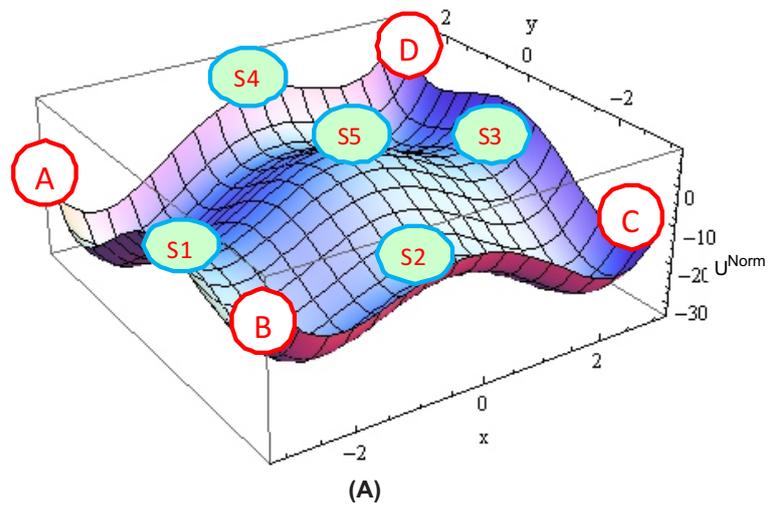

(A)

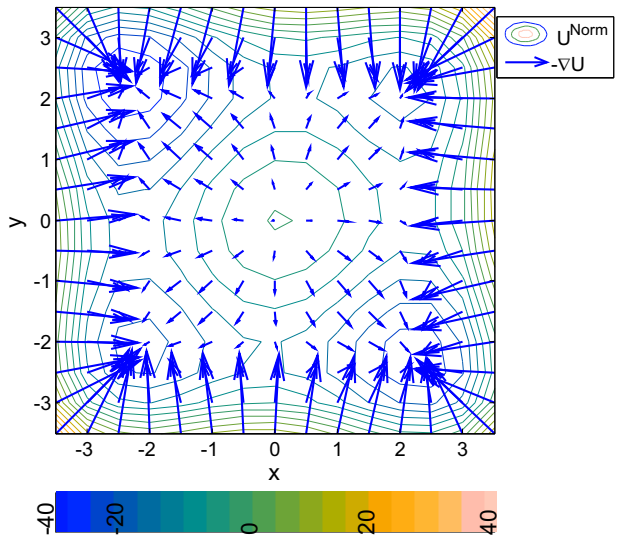

(B)

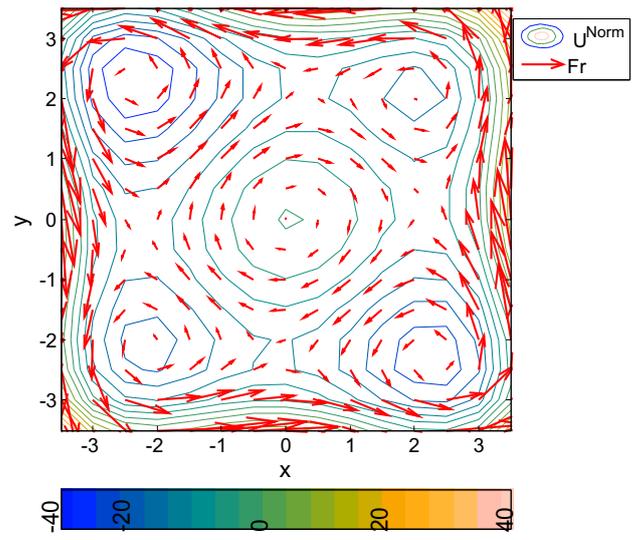

(C)

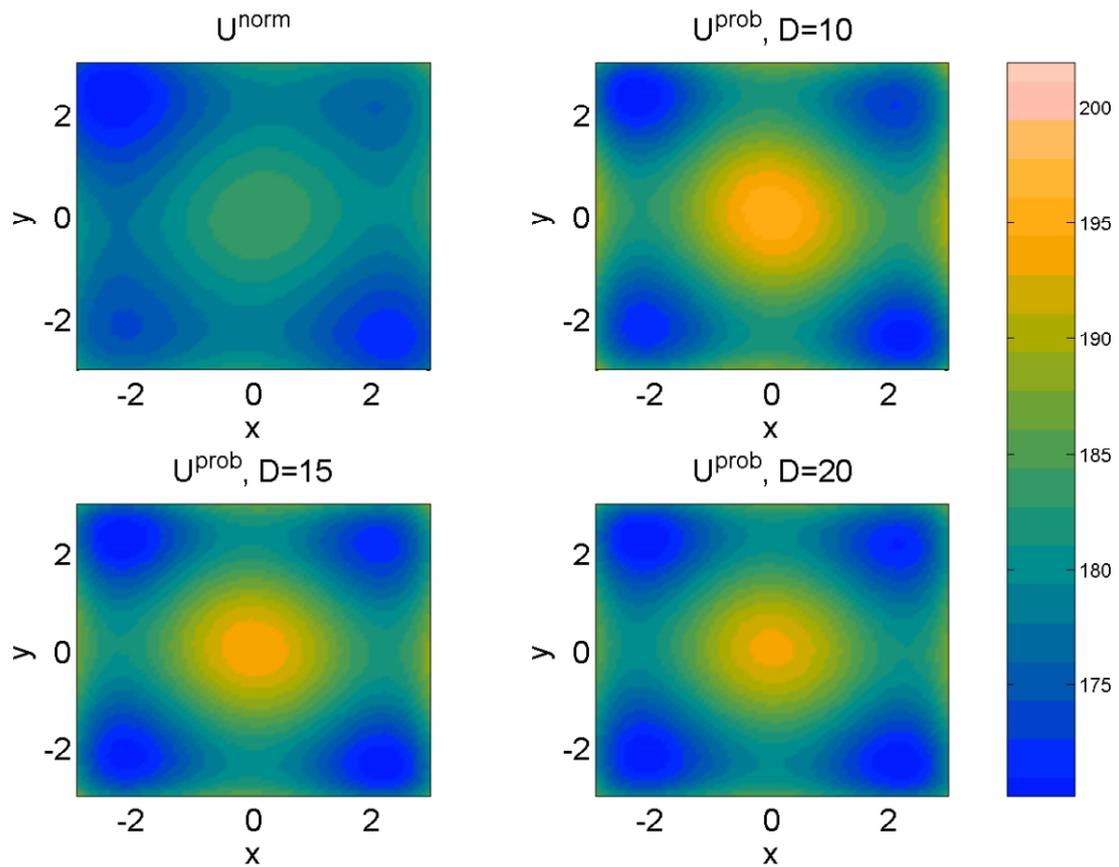

(D)

Fig. 4

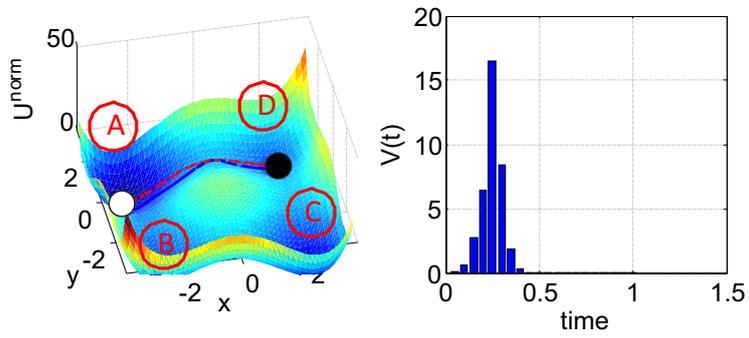
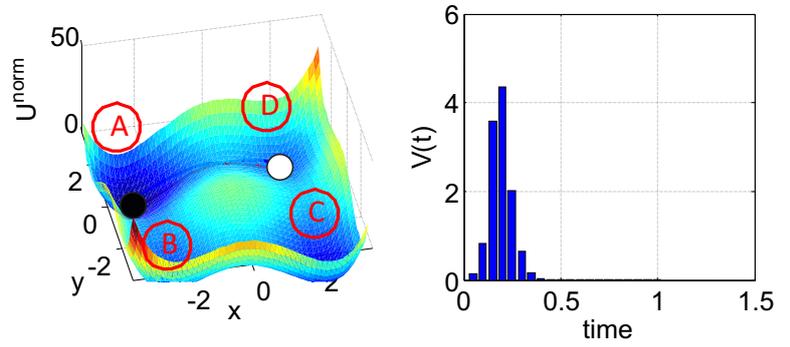
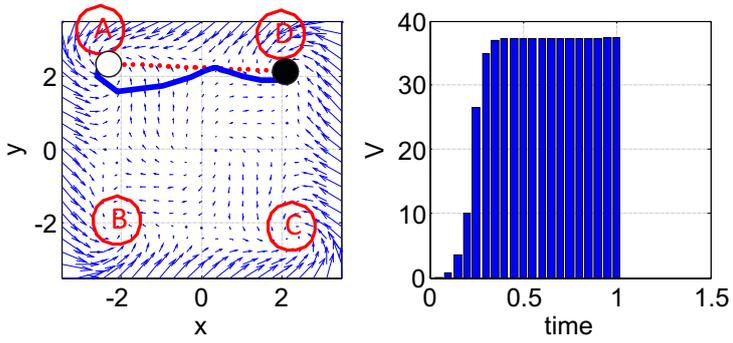
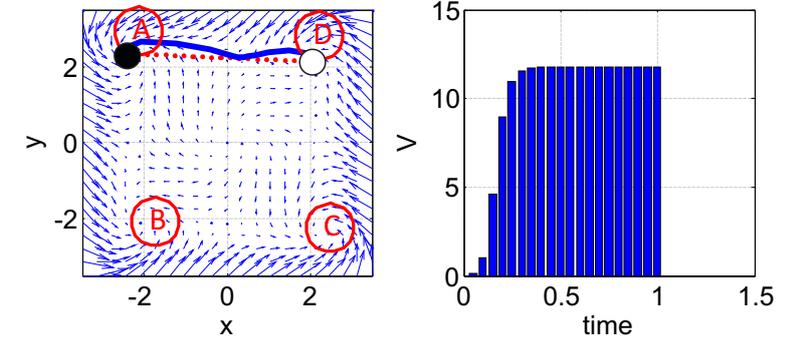

**(A)**  **(B)**

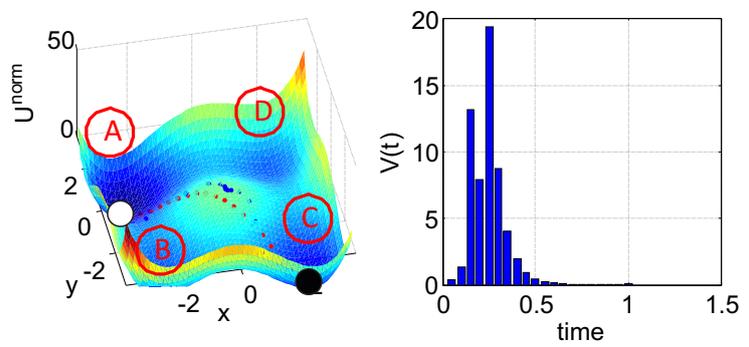
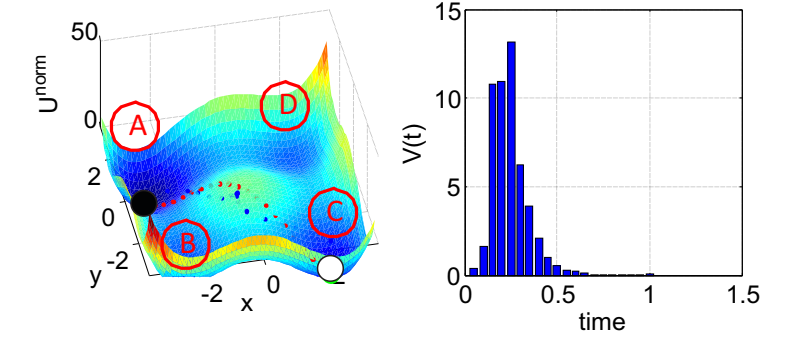
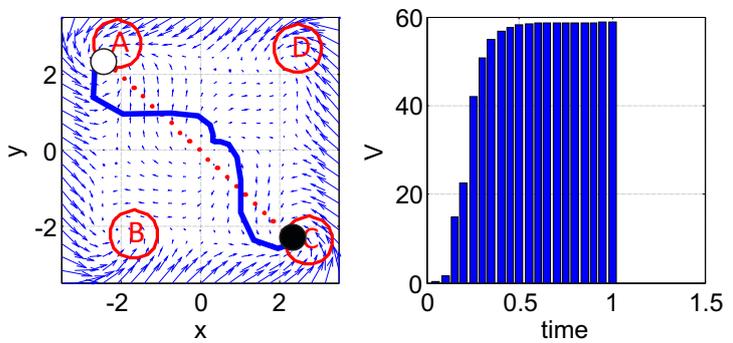
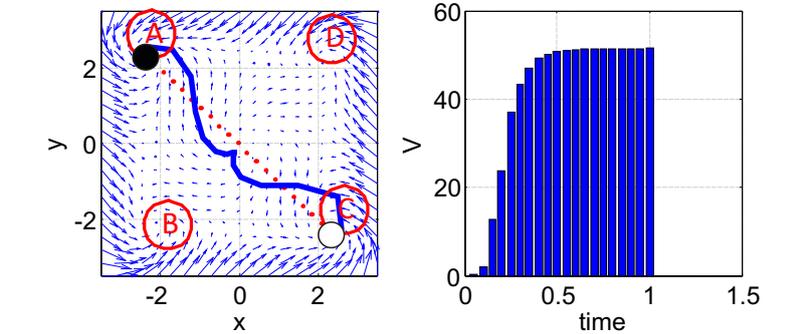

**(C)**  **(D)**

**Fig. 5**

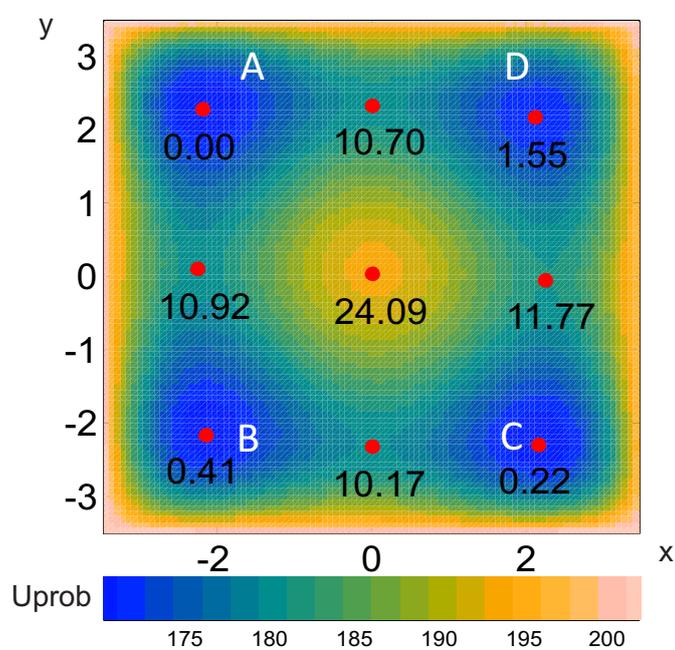
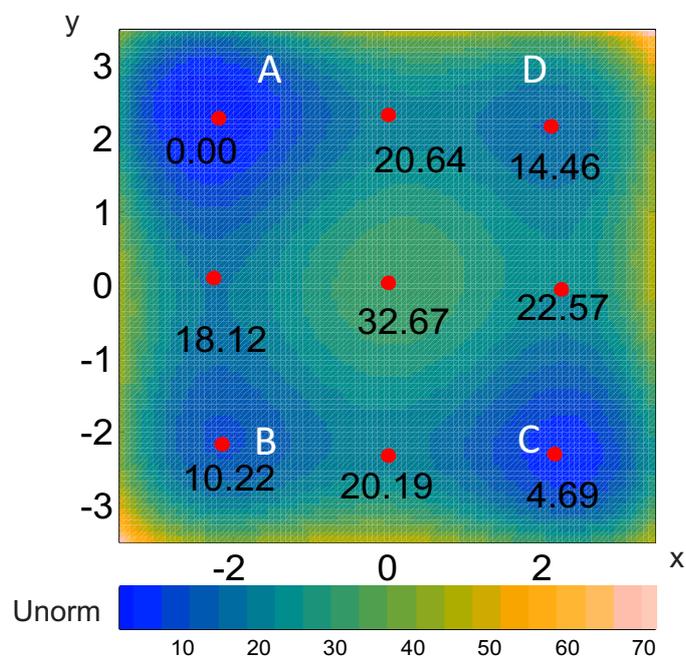

**(A)** **(B)**

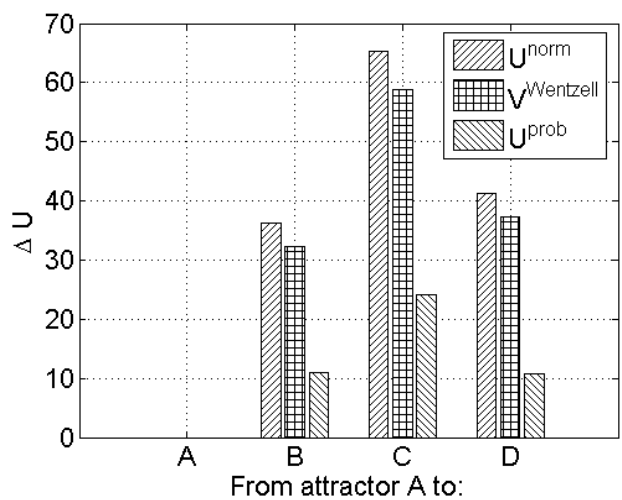
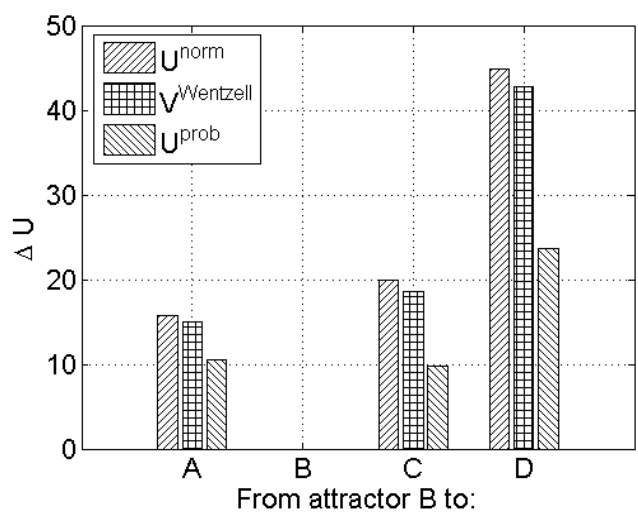
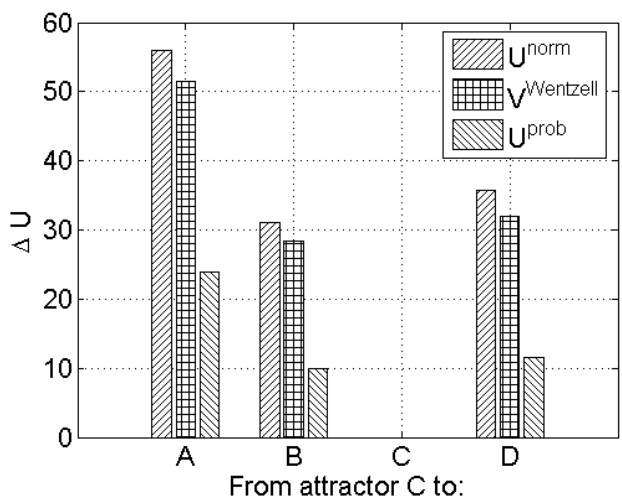
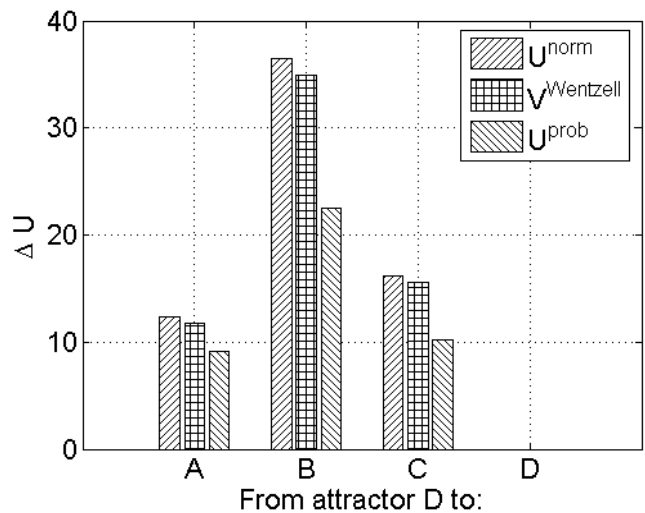

**(C)**

**Fig. 6**